\documentstyle[psfig,epsf]{mn}
\def\hone{\ifmmode{\rm H \/ {\sc i}}\else{H\/{\sc i}}\fi}
\def\rc{\ifmmode{r_{\!c}} \else{$r_{\!c}$} \fi}
\def\Rc{\ifmmode{R_{c}} \else{$R_{c}$} \fi}
\def\rth{\ifmmode{r_{tH}} \else{$r_{tH}$} \fi}
\def\Rtd{\ifmmode{R_{tD}} \else{$R_{tD}$} \fi}
\def\Md{\ifmmode{M_{\!d}} \else{$M_{\!d}$} \fi}
\def\dx{\ifmmode{\Delta x} \else{$\Delta x$} \fi}
\def\EJ{\ifmmode E_{\rm J} \else $E_{\rm J}$\fi}
\def\LS{L\&S}
\def\kpc{\ifmmode{{\rm \,kpc}} \else{${\rm \,kpc}$}\fi}
\def\x{$x$}
\def\y{$y$}
\def\vx{\ifmmode{v_{x}} \else{$v_{x}$} \fi}
\def\vy{\ifmmode{v_{y}} \else{$v_{y}$} \fi}
\def\lsun{L_{\odot}} %math mode only
\begin{document}

\title{The kinematics of lopsided galaxies}

\author[Edo Noordermeer, Linda S. Sparke and Stephen E. Levine]
{Edo Noordermeer$^{1, 2}$, Linda S. Sparke$^3$ and Stephen E. Levine$^4$\\
$^1$Astronomical Institute, Utrecht University, P.O. Box 80000, 3508 TA 
Utrecht, The Netherlands \\
$^2$Kapteyn Institute of Astronomy, P.O. Box 800, 9700 AV Groningen, 
The Netherlands, edo@astro.rug.nl (current address) \\
$^3$Department of Astronomy, University of Wisconsin, 475 N. Charter St., 
Madison, WI 53706 \\
$^4$US Naval Observatory, Flagstaff Station, P.O. Box 1149, Flagstaff, 
AZ 86002-1149}
\maketitle

\begin{abstract}

Lopsidedness is a common feature in galaxies, both in the distribution of 
light and in the kinematics. 
We investigate the kinematics of a model for lopsided galaxies that consists 
of a disc lying off-centre in a dark halo, and circling around the halo centre.
We search for families of stable, closed, non-crossing orbits, and assume that 
gas in our galaxies moves on these orbits.  
Several of our models show strong lopsided gas kinematics, especially the 
ones in which the disc spins around its axis in a retrograde sense compared 
to its motion around the halo centre. 
We are able to reproduce the \hone\ velocity map of the kinematically
lopsided galaxy NGC~4395. 

The lopsidedness in our models is 
%strongly correlated with the relative densities of the disc and the
%halo; it is 
most pronounced in the models where 
the halo provides a relatively large fraction of the total mass at small radii.
This %** supports the observed fact that 
may explain why the gas shows lopsidedness more frequently in 
late-type galaxies, which are dominated by dark matter. 
Surfaces of section show large regions of irregular orbits in the models where
the halo density is low. 
This may indicate that these models are unstable. 

\end{abstract}


\section{Introduction}
\label{sec:introduction}

It has been known for a long time that many disc galaxies are lopsided. 
Baldwin, Lynden-Bell \& Sancisi \shortcite{Baldwin} were among the first to 
note that many galaxies show large-scale asymmetries in the optical images and 
the gas distributions. 
Richter \& Sancisi \shortcite{RS} examined a large sample of 
about 1700 global \hone-profiles, and concluded that at least 50\% of all disc 
galaxies show significant asymmetries. This result has recently been 
confirmed by Haynes et al.\ \shortcite{Haynes}, who studied 104 \hone-profiles 
of spiral galaxies. Matthews, Van Driel \& Gallagher \shortcite{Matthews1} 
studied \hone-profiles of 30 extreme late-type spiral galaxies. These authors 
found that about 75\% of the galaxies in their sample have more or less 
asymmetric profiles, and inferred that lopsidedness is more common
among these very late-type spirals. 

Swaters \shortcite{Swatersthesis} studied a sample of 75 late-type galaxies in 
\hone\ and red light, finding a high incidence of lopsidedness in both the
distribution of gas and its kinematics.
Swaters et al.\ \shortcite{SwatersMN} presented global profiles,
rotation curves and velocity fields for two of these, 
DDO~9 and NGC~4395 (of type Im and Sm respectively), which are both 
kinematically lopsided, although neither has nearby companions.  
In these and other kinematically lopsided galaxies, the rotation curves level 
off at a constant velocity on one side of the galaxy, while they keep on 
rising on the other side. The contours of the velocity fields are more 
strongly curved on the side with the flat rotation curve, and the
\hone-profile is asymmetric too.
 
Many galaxies are also lopsided in their light distribution. 
Rix \& Zaritsky \shortcite[hereafter R\&Z]{RZ} studied K-band images
of 18 disc galaxies and found that about one third of them were
significantly lopsided. R\&Z argue that lopsidedness should be more
common in the \hone-distribution than in the stars, because of the
longer dynamical timescales in the outer part of the disc.
Zaritsky \& Rix \shortcite{ZR} followed up with I-band
imaging of a larger sample of 60 galaxies, finding $\sim 30\% $ of
them to be lopsided.  Rudnick \& Rix \shortcite{RR} explored
earlier-type systems of type S0 to Sab, and found roughly 20\% of
lopsided disks.

The relation between lopsidedness in the distribution of starlight or
gas and an asymmetric velocity field is not at all clear. NGC 891
shows a strongly lopsided gas distribution, with gas extending much 
further from the center on the southern than on the northern side. But 
up to the radius where the \hone\ stops on the northern side, the
kinematics are almost perfectly symmetric with no obvious deviations 
from circular motion; beyond that, the rotation curve on the south
side remains flat \cite[Fig.\ 2]{SA,SSvdH}. 
By contrast, both the stellar light and the
\hone\ distribution in DDO~9 and NGC~4395 are fairly symmetric, but
the velocity fields are quite lopsided \cite{SwatersMN}. 
Kornreich et al. \shortcite{Kornreich1,Kornreich2}
compare morphological and kinematical lopsidedness in two
samples, each of nine spiral galaxies, and find little or no
correlation between the different parameters they measured. 
In the latter paper, they state that `normal morphology is not an
indicator of normal kinematics and, conversely, that perturbed
kinematics do not necessarily manifest as perturbed optical
morphology'. Swaters \shortcite{Swatersthesis} looked at
the \hone-distribution and kinematics of 73 late-type dwarf galaxies,
and found 19 galaxies with stronger kinematical than morphological
asymmetries, 18 where the lopsidedness is stronger in the density, and
5 with comparable asymmetries in both (Chapter~3, Table~A2).

The origin and persistence of lopsidedness in galaxies have not yet
been explained in a satisfactory way. In both the radio and near-infrared
surveys, the asymmetries affect large parts of the discs; they are not small, 
localized irregularities.  
The fact that so many galaxies, including isolated systems, are lopsided 
argues against transient effects. 
Lopsidedness may be a general and intrinsic feature of disc galaxies. 

Baldwin et al.\ \shortcite{Baldwin} suggested a model in which the gas
and stars of a lopsided disc move in an axisymmetric potential on a
series of initially aligned elliptical orbits.  Due to differential
precession, this lopsidedness will wind up into a single leading
spiral arm and slowly disappear.  These authors estimate that the
asymmetry in their model can persist for about 5 Gyr, but claim that
this lifetime is too short to explain the observed frequency of
lopsidedness in nature.  Another drawback is that, since the orbiting
gas conserves its angular momentum, the inferred rotation curve must
be significantly higher on the `short' side of the disk.

The only stellar-dynamical models of disc galaxies that develop strong
lopsided asymmetries are models with many counter-rotating particles
\cite{ZH,Sawamura,SM}.  However, counter-rotation is only rarely
observed in disc galaxies (see e.g. Kuijken, Fisher \&
Merrifield 1996).  Syer \& Tremaine \shortcite{ST} present fluid models for
thin, scale-free, lopsided disks; their solutions are gravitationally
self-consistent, and the distortion is stationary in inertial space.
Galli et al.\ \shortcite{Galli} show that these configurations are secularly
stable.
In these disks, a lopsided surface density implies lopsided
kinematics.  Earn \& Lynden-Bell \shortcite{ELB} reach a similar conclusion in
their work on self-gravitating lopsided disks: the velocity field of a
lopsided disk should itself be asymmetric.  Such a model cannot
account for the symmetric rotation curves of systems such as NGC~891. 

Zaritsky \& Rix \shortcite[hereafter Z\&R]{ZR} tried to explain lopsidedness 
as a result of recent accretion of small satellite galaxies on the parent 
galaxy. Using results of N-body simulations by Walker, Mihos \& Hernquist 
\shortcite{Walker}, Z\&R conclude that lopsidedness can be caused by accretion 
of a satellite within the last 1~Gyr. Using this lifetime and the observed 
frequency of lopsidedness, they obtain an upper limit on the accretion rate. 
However, the accretion rate can be much lower if the lopsidedness is more 
persistent, or if there are other mechanisms responsible for it. In particular,
R\&Z point out that the winding problem is avoided if the potential of the 
galaxy is lopsided itself, with the gas and stars just responding; they do not 
address the question of how to get a lopsided potential in the first place. 

Recently, attention has been drawn to models in which the dark halo
accounts for an asymmetry in the overall galaxy potential. 
Weinberg and collaborators \cite{Weinberg1,Weinberg2,VW} have shown that
lopsided modes of oscillation in a spherical system may have very
long decay times, so that asymmetries in the galactic halo can
persist long after the original disturbance.  

Jog \shortcite{Jogone} showed that even a small lopsided perturbation
of the galaxy halo can cause strong asymmetry in the density of disk
gas, in a sense opposing the perturbation in the halo. Investigating the
self-consistent response of the disk, Jog \shortcite{Jogtwo} showed
that the disk's gravity indeed opposes the lopsidedness of the halo.
As a result, the disk becomes strongly lopsided only in its outer
parts, where the halo is dynamically dominant. 
However, Jog does not show model rotation curves.

Levine \& Sparke \shortcite[hereafter \LS]{LS} presented a model to
explain lopsidedness, with the disc lying off-centre in the dark halo
and orbiting around its centre.  This model has some aspects in common
with that presented by de Vaucouleurs \& Freeman \shortcite{VF} for
off-centre bars in galaxies such as the Large Magellanic Cloud.   
The N-body simulations
by \LS\ showed that when the rotation of the disc is retrograde with
respect to its motion around the centre of the halo, the disc can
remain off-centre for many rotation periods.  The retrograde disc
appears to settle into a circular orbit close to the core radius of
the halo, inside which the halo density is roughly constant; a
prograde disc drifted slowly inward towards the centre of the halo.

This paper will explore particle orbits in the \LS\ model; we
investigate the kinematics of gas moving in such a model galaxy. 
We explain our model potential in section~\ref{sec:model}.
In section~\ref{sec:results}, we present surfaces of section,
rotation curves, and velocity fields.  We show that orbits in this model
can account both for disks with a lopsided surface density but
symmetric kinematics, and for objects like NGC~4395, in which the
velocity field is more distorted than the surface density.  
We also explore the effects of changing the basic
model parameters.  In section~\ref{sec:discussion} we present a
discussion of our conclusions.

\section{Kinematics of a lopsided model}
\label{sec:model}

\LS\ presented N-body simulations of a model for lopsided galaxies, 
consisting of a disc of particles rotating around the centre of a massive 
dark halo, represented by a fixed spherical potential. 
Here, we compute stable, closed, non-crossing periodic orbits in an analytical 
representation of the potential of this model. 
Assuming that the gas in the disc moves on those orbits, we find the resulting 
velocity-field and rotation curves. 
We expect the disc particles to follow paths that oscillate about
these stable periodic orbits, and use surfaces of section to explore
the regions of trapped orbits.  
Although \LS\ found that the retrograde-spinning disc remained off-centre
in the halo longer than the prograde disc, we consider both cases.

\subsection{A model for the gravitational potential}
\label{subsec:model}

Our model consists of a flat disc and a spherical halo. We confine 
our attention to orbits in the plane of the disc. The disc is an 
axisymmetric Kuzmin-Toomre disc of mass $M_{KT}$, lying in the \x-\y\ plane. 
Its potential at distance $R$ from the disc centre in that plane is given by
\begin{equation}
\Phi_{KT}(R) = - \frac  {  G M_{KT}  }{  \sqrt{R^{2} + R_{c}^{2}}  }\, , 
\label{eq:discpot}
\end{equation}
where G is the gravitational constant, and \Rc the scale length of the 
disc. We define our units such that $G=1$ and $\Rc = 1$. In the 
simulations of \LS, the disc was truncated at radius $\Rtd = 10$. Its
total mass is then given by: 
\begin{equation}
M\left(<R_{tD}\right) = M_{KT} \left( 1 - \frac {1} 
			{\sqrt{1 + R_{tD}^{2}/R_{c}^{2} } } \right) \, .
\label{eq:discmass}
\end{equation}

The halo is spherically symmetric and has a pseudo-isothermal form. 
The density at distance $r$ from its centre is given by
\begin{equation}
\rho_{H}(r) = \frac{\rho_{c}} {1 + (r/r_{c})^{2}} \, ,
\label{eq:halodensity}
\end{equation}
where $\rho_{c}$ is the central density and $r_{c}$ the core radius of the 
halo, the radius where the density falls to half the central value. 
The velocity $V_{H}$ of a circular orbit at radius $r$ in this halo is given 
by:
\begin{equation}
V_{H}^{2}(r) = V_{\infty}^{2} \left[ 1 - \frac{\rc}{r} 
	\arctan \left( \frac{r}{\rc} \right) \right] \, .
\label{eq:halov}
\end{equation}
The circular velocity rises linearly in the centre, and approaches the 
asymptotic velocity $V_{\infty}$ at large radii:
\begin{equation}
V_{\infty}^{2} = 4 \pi G \rho_{c} r_{c}^{2} \, .
\label{eq:Vinf}
\end{equation}
The halo described by equation~\ref {eq:halodensity} has infinite mass, but 
\LS\ truncated the halo at $\rth = 20$. 
Its total mass is then given by:
\begin{equation}
M \left( <r_{tH} \right) = \frac{ V_{\infty}^{2} }{ G } 
	\left[ \rth - \rc \arctan \left(\frac{\rth}{\rc}\right) \right]  \, .
\label{eq:halomass}
\end{equation}

We define our unit of mass such that the total mass in the truncated model 
(disc + halo) is unity. 
We define $\Md$  as the fraction of the total mass that is in the truncated 
disc ($0 \leq \Md \leq 1$). 
The mass of the truncated halo is then $1 - \Md$. 
We also need to know the contribution from the halo to the mass in the inner 
regions of the disc. 
We define $\eta$ as the fraction of the total mass within a sphere of radius 
$r=2$ from the disc centre that belongs to the halo.
While the halo as a whole is much more massive than the disc, the disc 
dominates within these inner parts.   

For $\Rc = 2 \kpc$ and a total mass of $10^{11} {\rm M_{\odot}}$, our time 
unit corresponds to $4.22{\rm \,Myr}$ and our velocity unit to $464 
{\rm \, km \, s^{-1}}$. 

The halo is centered at the origin, while the disc centre is placed at a 
distance \dx on the positive \x-axis. 
The combined potential in the \x-\y\ plane is then given by:
\begin{equation}
\Phi_{tot}(x,y) = \Phi_{H}(x,y) + \Phi_{KT}(x-\dx, y) \, , 
\label{eq:totalpot}
\end{equation}
where $\Phi_{H}$ is the halo potential.
The halo is much more massive than the disc, so we let the whole system rotate 
around the origin at an angular velocity $\Omega_{\! P}$ equal to the circular 
rotation velocity in the halo potential at the radius \dx of the disc centre:
\begin{equation}
\Omega_{\! P}^{2} = \frac{V_{\infty}^{2}}{\dx^{2}}  
	\left[ 1 - \frac{r_{c}}{\dx} \arctan \left( \frac{\dx}{r_{c}} 
							\right) \right] \, .  
	\label{eq:Omega}
\end{equation}

Now, we look for orbits that are closed in the corotating frame. 
Defining the angular momentum vector ${\bf \Omega} = \Omega_{\! P} 
{\bf \hat{e}_{z}}$, the equation of motion in this frame is:
\begin{equation}
{\bf \ddot{r} } = - \nabla \Phi_{tot} - 2 ({\bf \Omega \times \dot{r} }) - 
		{\bf \Omega \times ( \Omega \times r) } \, ,    
		\label{eq:eqofmotion}
\end{equation}
where the second and third term on the right side are the Coriolis and 
centrifugal forces respectively. The effective potential $\Phi_{e\!f\!f} 
\equiv \Phi_{tot} - \frac{1}{2} \Omega_{\! P}^{2} \left( x^2 + y^2 \right)$ is 
defined as the sum of the gravitational potential and a `centrifugal 
potential'. 
Orbits in this system admit an isolating integral of motion, the Jacobi 
integral:
\begin{equation}
\EJ = \frac{1}{2} |{\bf \dot{r}}|^{2} + \Phi_{e\!f\!f} 
\label{eq:Jacobi's}
\end{equation} 
(see Binney \& Tremaine 1987, section 3.3.2). % Not a moving link!!

\subsection{Finding orbits for gas flow}
\label{subsec:method}

We assume that gas in our model moves in stable, closed and non-crossing 
orbits. The flow is then smooth and laminar, without major shock fronts. 
We used the method of surfaces of section to find closed orbits (see e.g. 
Binney \& Tremaine 1987, sections 3.2 to 3.5). % Not a moving link!!
For a range of values of \EJ, we integrated orbits with different initial 
conditions, such that we probe the entire phase-space. 
Each time the orbit crosses the \x-axis with $\vy > 0$, we mark its 
\x-coordinate and \x-velocity in an $x\dot{x}$-diagram. 
The simple closed orbits that we seek always cross the \x-axis at the same 
point, with the same velocity; they will appear in the diagram as single 
points. 

We then check if the orbits are stable, using a method first proposed by 
H\'{e}non \shortcite{Henon}. 
Also, we check that orbits do not cross each other. 
We can then imagine filling with gas all the stable orbits that do not 
cross each other, to make a model velocity field or a rotation curve. 
Gas would not generally be found on unstable orbits or orbits that cross each 
other, so we ignored those orbits.
However, stars could still be trapped around stable crossing orbits.

For the integrations, we used a variable-order, variable-step Adams method, 
implemented in Fortran77. 
This method maintains high accuracy during the long integrations needed to 
make the surfaces of section. 
We made use of subroutines from the NAG-library, using double precision 
variables, and a tolerance of $10^{-10}$; a smaller tolerance slowed down the
calculations and did not improve the accuracy of the integrations.

\section{Model rotation curves and velocity fields}
\label{sec:results}

The parameters of our seven models are given in table~\ref{table:parameters}. 
The disc scale length $\Rc$ is 1 by definition, and the disc 
and halo truncation radii $\Rtd$ and $\rth$ are 10 and 20 respectively. 
\begin{table}
 \begin{center}
  \begin{tabular}{c|ccccccc} \hline
	Model & \Md    & \Rc  & \rc  & \dx  & \Rtd & \rth & $\eta$ \\ \hline
	A     & 0.092  & 1    & 2    & 2.5  & 10   & 20	  & 19\%   \\
	B     & 0.092  & 1    & 2    & 1.5  & 10   & 20	  & 25\%   \\
	C     & 0.092  & 1    & 2    & 5    & 10   & 20	  & 8\%    \\
	D     & 0.186  & 1    & 2    & 2.5  & 10   & 20	  & 9\%    \\
	E     & 0.046  & 1    & 2    & 2.5  & 10   & 20	  & 33\%   \\
	F     & 0.092  & 1    & 1    & 1.5  & 10   & 20	  & 37\%   \\
	G     & 0.092  & 1    & 4    & 4    & 10   & 20	  & 8\%    \\ \hline
  \end{tabular}
 \end{center}
\caption{Values of parameters for the different models. \Md is the fraction 
  of the total mass in the truncated disc; \Rc is the disc scale length; \rc 
  is the halo core radius; \dx is the distance between halo and disc centre; 
  \Rtd is the truncation radius of the disc; \rth is the truncation radius of 
  the halo; $\eta$ is the contribution from the halo to the total mass
  within a sphere of $r=2$ around the disc centre.}
\label{table:parameters}
\end{table}
For all models, we consider both orbits that circulate the disc centre in a 
prograde and retrograde sense with respect to the disc's motion around the halo
centre. 

\subsection{The starting model}
\label{subsec:Model A}

\begin{figure*}
  \epsfbox{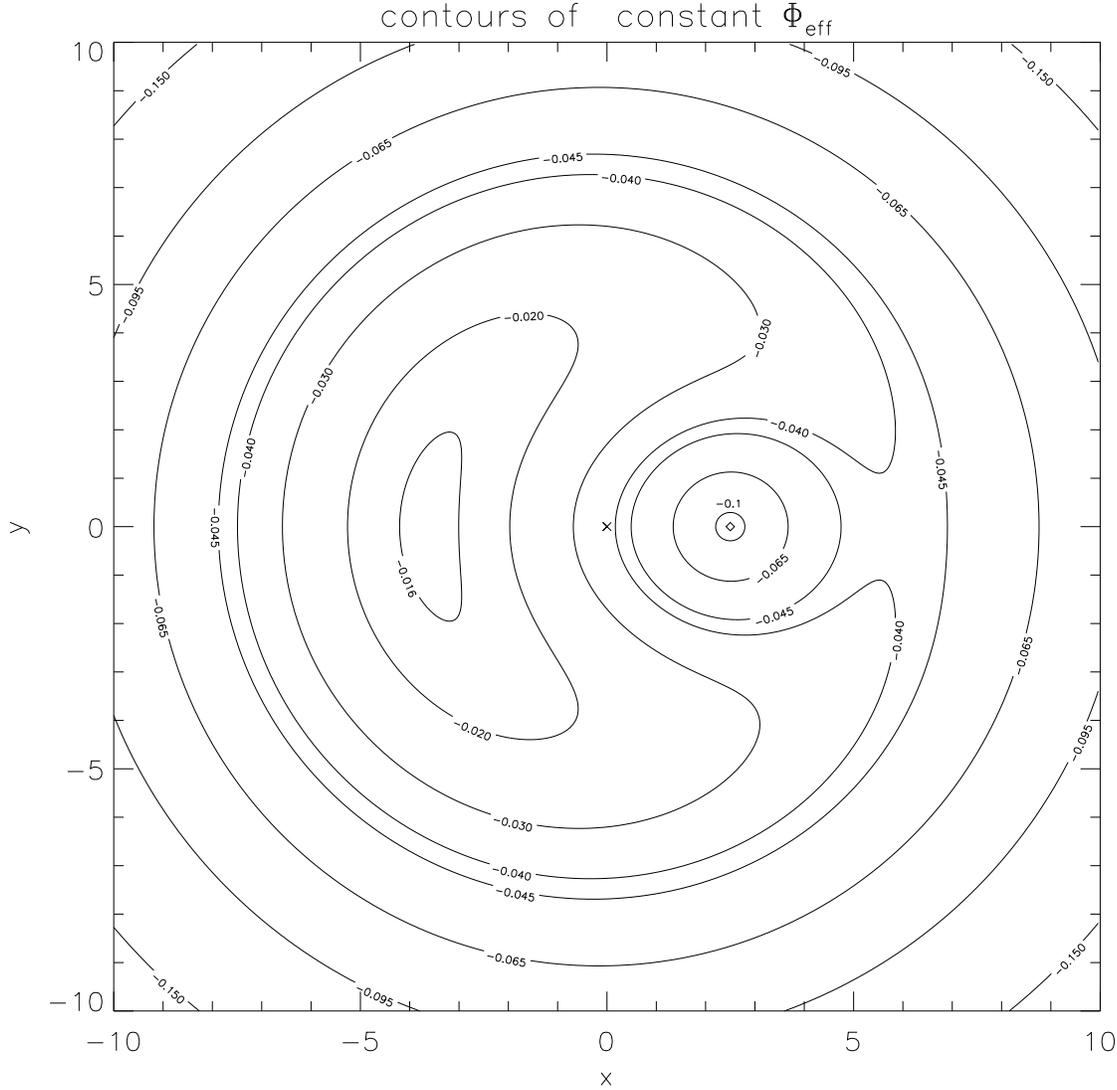}
  \caption{contours of constant $\Phi_{e\!f\!f}$ for model A. The disc centre
   lies at $(2.5,0)$ and is marked by the diamond; the entire system is 
   rotating counterclockwise around the origin, marked by the cross.}
  \label{fig:Veff}
\end{figure*}

Model~A was inspired by one of the most successful runs in \LS. 
They found that a retrograde spinning disc sank towards the halo centre until 
the most tightly-bound particles of the disc orbited at a radius close to the 
halo core radius. 
We chose $\Md = 0.092$, $\rc = 2$ and $\dx = 2.5$, 
so that the disc centre lies just outside the halo core. 
The halo contributes 19\% of the total mass within a distance $r=2$ from the 
disc centre. Fig.~\ref{fig:Veff} shows a contour-plot of the effective 
potential $\Phi_{e\!f\!f}$ in the co-rotating coordinates. 
In the region near $(2.5, 0)$, the potential is clearly dominated by the disc. 

\begin{figure*}
  \epsfbox{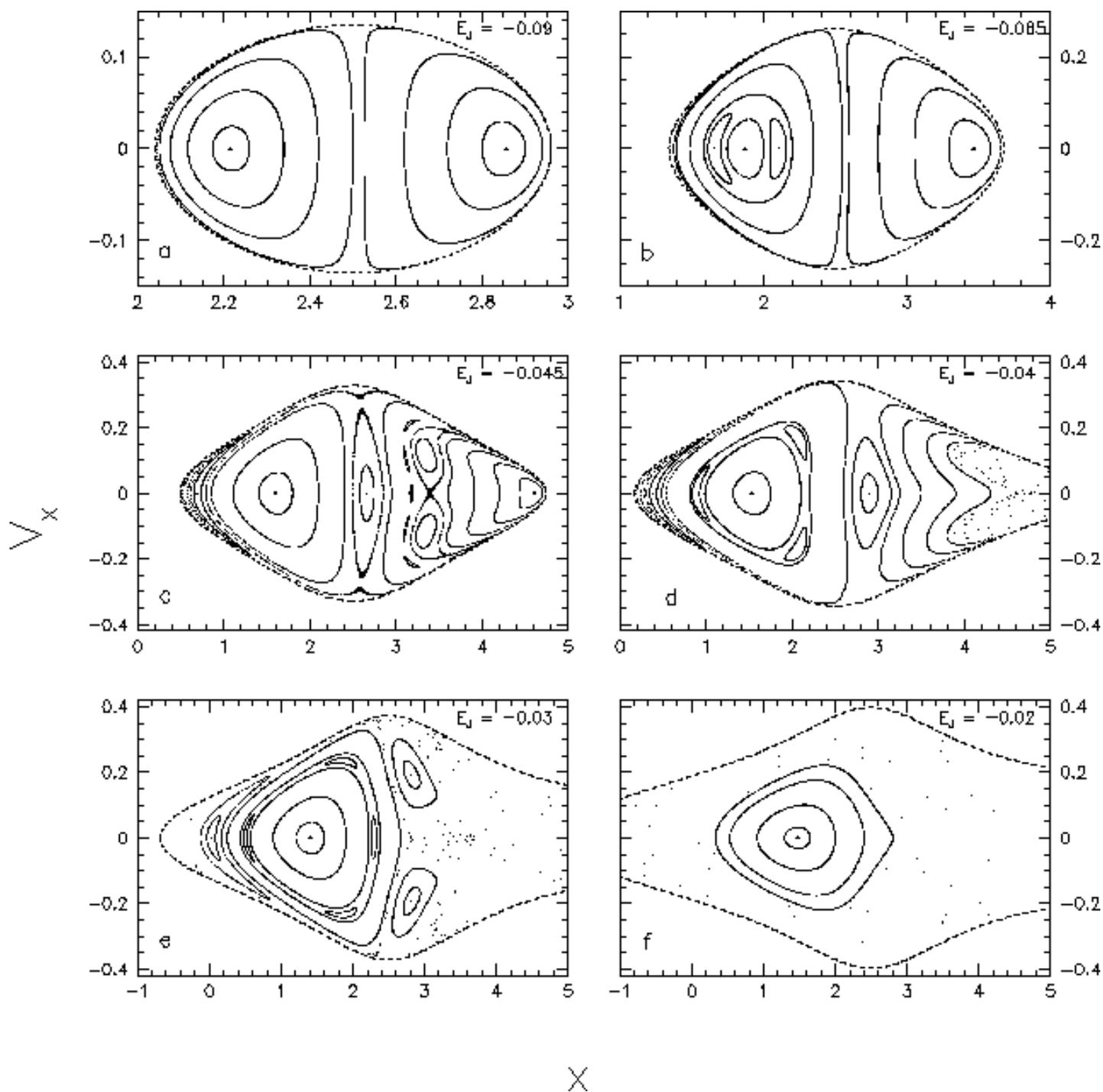}
  \caption{Surfaces of section for $y=0$ and $\vy\geq0$ for model A. Note that
   the scales differ between the panels. The value of the Jacobi integral is 
   indicated in each panel. The dashed curve is the zero velocity curve, where
   $\vy=0$. Simple closed orbits are indicated by triangles. All coordinates
   are measured in a frame corotating with the disc centre.}
  \label{fig:surfsecsA}
\end{figure*}

When the Jacobi integral is low, the surfaces of section in 
Fig.~\ref{fig:surfsecsA} show two simple closed orbits.
One crosses the \x-axis at $x > 2.5$, and circles the disc centre in a 
prograde sense with respect to the motion of the disc around the halo centre.
The other is a retrograde orbit, and crosses the \x-axis at $x < 2.5$. 
Both these orbits are stable and surrounded by a family of non-closed but 
regular orbits. 
Those appear as the concentric curves, nested around the closed orbits. 

As the Jacobi integral is increased, most of the retrograde orbits remain 
regular, and regular orbits trapped about the simple closed orbit continue to 
occupy a large area to the left. 
There are some `islands' of orbits that close after multiple circuits; see for
example in panel b, near $(1.6, 0)$ and $(2.1, 0)$, an orbit that closes after
two circuits. These do not occupy a large region in phase space. 
The structure of the prograde orbits becomes quite complicated. 
We see many small `islands' of multi-periodic orbits and we also find 
irregular or chaotic orbits; they fill a two-dimensional area in the diagrams 
and produce `fuzzy' regions such as that to the right of $x=4$ in panel d. 
For $\EJ = -0.045$, the closed prograde orbit has shifted all the way to the 
edge of the surface of section; for $\EJ = -0.04$ and higher, there is no 
closed, simply-periodic prograde orbit at all. 

\begin{figure*}
  \epsfbox{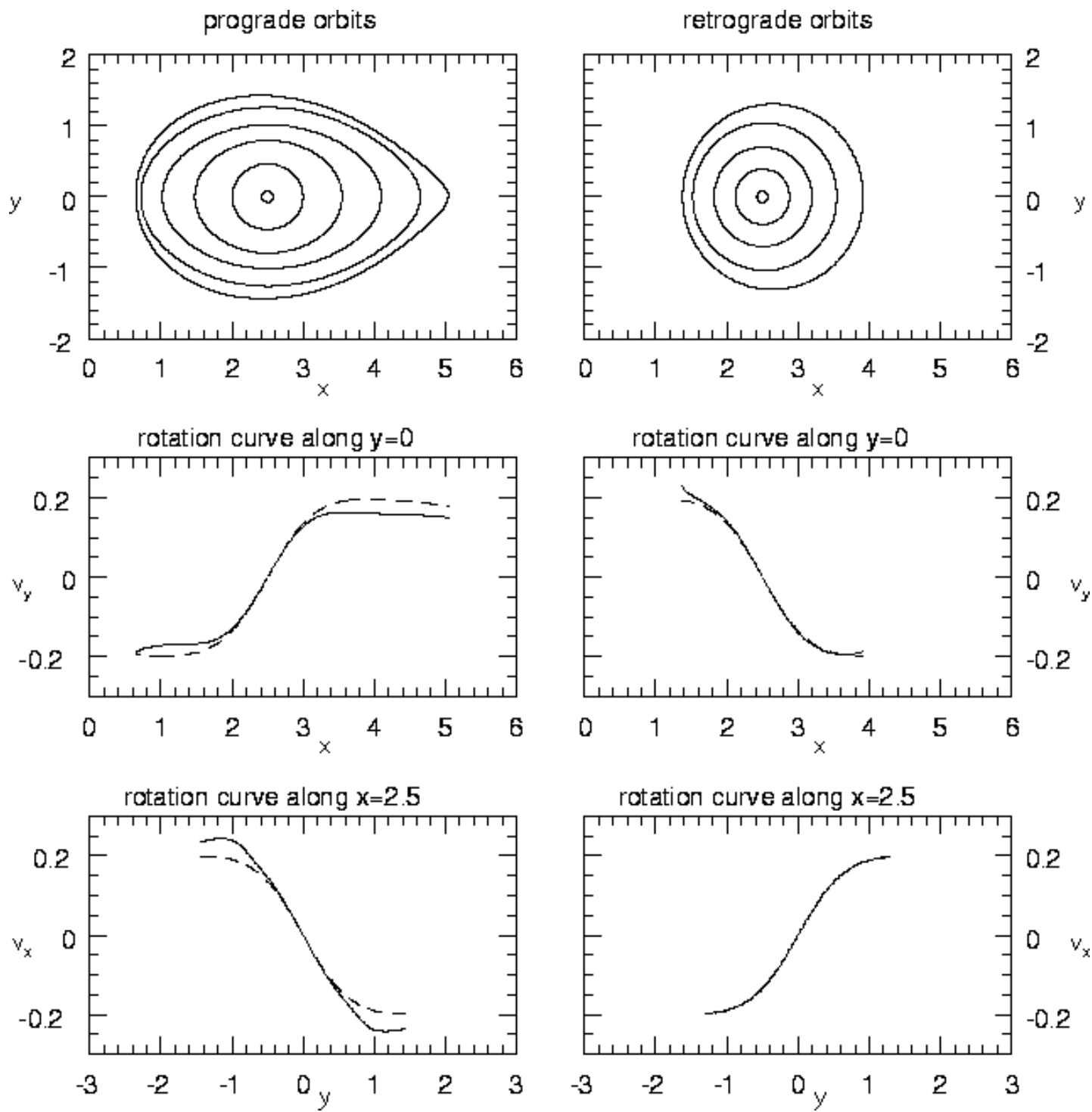}
  \caption{The upper panels show closed orbits as seen in the corotating frame 
   	for model A. The middle panels show the rotation curves along the 
	\x-axis. The velocities are measured in an inertial frame, relative to 
	the disc center. The dashed curves show the theoretical rotation curve
	for an isolated Kuzmin-Toomre disc of the same mass, without the halo
	present. The lower panels show the symmetric rotation curves along the
	axis $x=2.5$. The left panels are for the prograde orbits, the right 
	ones for the retrograde orbits.}
  \label{fig:orb+curA}
\end{figure*}

When $\EJ \geq -0.04$, the zero velocity curve is not closed, so orbits are no 
longer bound to the disc and particles could in principle escape from the 
system. 
But regular retrograde orbits keep occupying a substantial region of phase 
space, until the Jacobi integral rises slightly above $-0.02$, when no regular 
orbits exist in the disc anymore.
This helps to explain why the retrograde-spinning disc could remain stable in
an orbit just outside the halo core in the simulations of \LS; 
even at high values of $\EJ$, many regular orbits trapped around the closed 
periodic orbits were available to the simulation particles.
In the prograde disc, although closed orbits persist to larger radii than in 
the retrograde case (see Fig.~\ref{fig:orb+curA}), even at $\EJ = -0.045$ 
(panel c) only a small part of the phase space is occupied by orbits trapped 
around them.
Thus it would be much more difficult to populate the prograde disc
self-consistently with simulation particles.

In the top panels in Fig.~\ref{fig:orb+curA}, we show the families of stable 
simply-closed orbits, both prograde and retrograde, as seen in the frame 
corotating with the disc, out to the radius where the closed orbits begin to 
cross each other.
Near the disc centre, where the potential of the system is dominated by the 
axisymmetric potential of the disc itself, the orbits are nearly circular. 
At larger radii, the prograde orbits become elongated in the $x$-direction. 
The retrograde orbits remain quite circular, but the geometric centre shifts 
to larger \x-values, away from the halo centre. 

In the middle panels, we plot the rotation curves along the \x-axis, as they
would be measured by an inertial observer, relative to the velocity of the 
centre of the disc. The dashed curves correspond to circular orbits in the disc
alone, without the halo present. 
For the prograde orbits, the rotation curve is fairly symmetric, but it
always lies below the one in the isolated disc; if the mass of the
galaxy were calculated from this rotation curve, it would be under-estimated. 
The retrograde curve shows the interesting feature that on the side closer to 
the halo centre, it rises all the way to the last non-intersecting
orbit, while on the other side it becomes flat.
The expected shape of the global profiles in \hone\ will depend on how much gas 
is present on these orbits. Generically, the peak on the
side where the rotation curve is flat will be higher, while the side with the 
lower peak has a sloping `shoulder' to high velocity.

In the panels at the bottom, we plot the rotation curves along the minor axis 
of the system, perpendicular to the axis of symmetry. These curves are exactly 
symmetric. The prograde curve now rises significantly above the one for an 
isolated disc, while the retrograde curve is almost indistinguishable from it.

\begin{figure*}
  \epsfbox{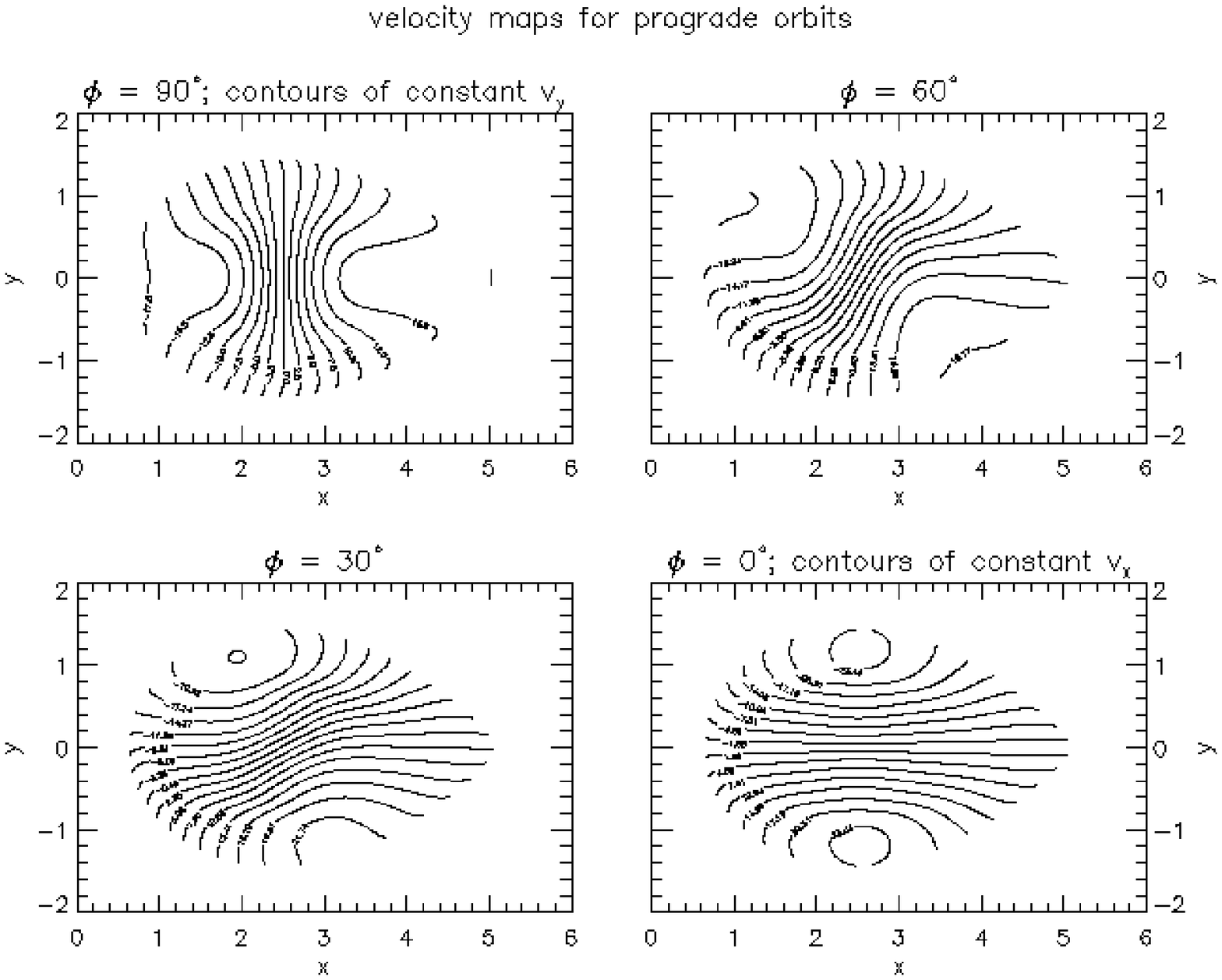}
  \caption{Contours of constant velocity along the prograde orbits of model A.
    The velocities are measured in an inertial frame, relative to the disc 
    centre. The directions of the velocities are measured by $\phi$, the angle 
    between the velocity and the \x-axis. The labels on the contours are in 
    units of $0.01$.}
  \label{fig:velmapproA}
\end{figure*}

\begin{figure*}
  \epsfbox{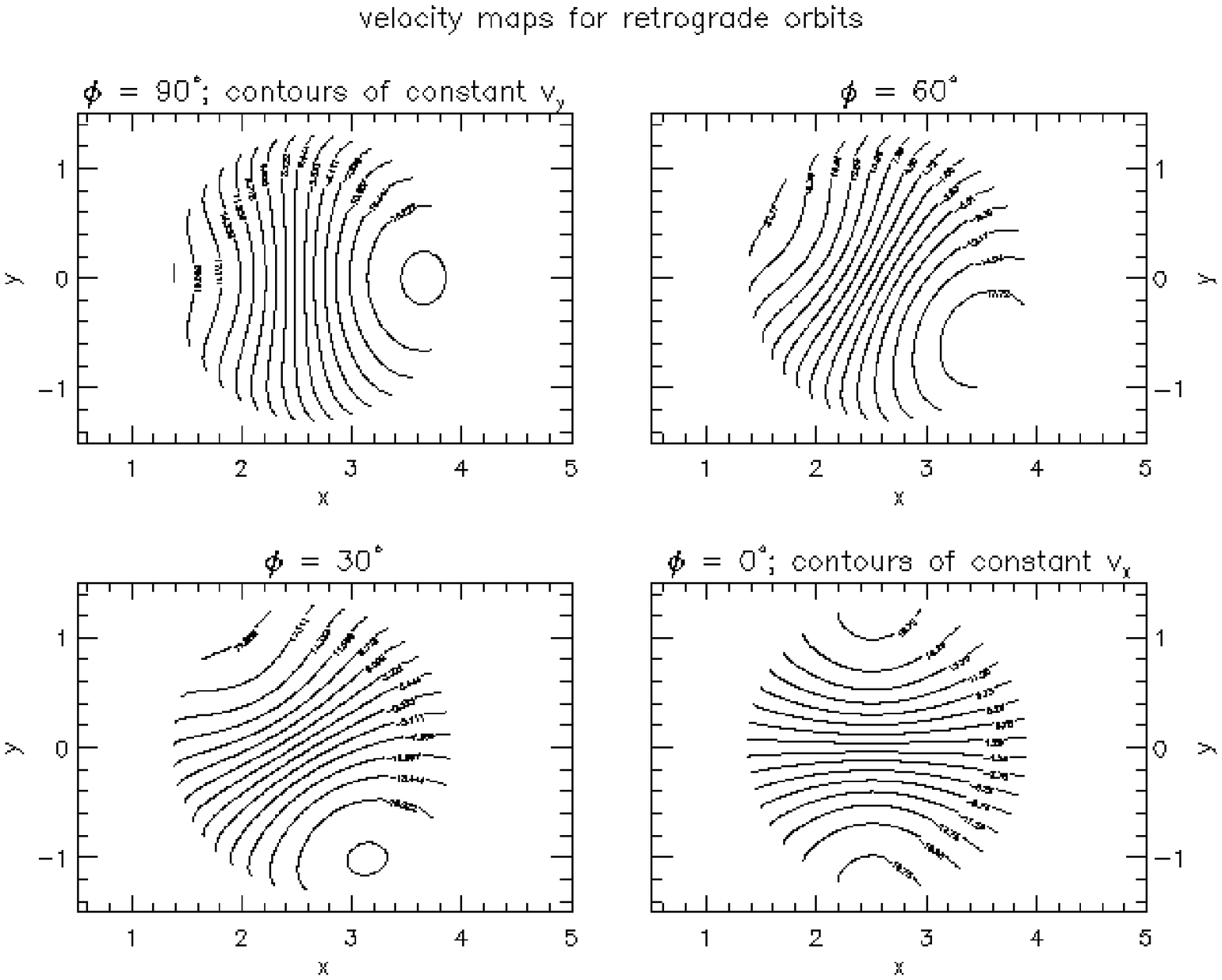}
  \caption{Same as Fig.~\ref{fig:velmapproA}, now for the retrograde orbits of 
    model A. Note that the scales along the axes are different from 
    Fig.~\ref{fig:orb+curA} and \ref{fig:velmapproA}. }
  \label{fig:velmapretroA}
\end{figure*}

In Fig.~\ref{fig:velmapproA} and \ref{fig:velmapretroA}, we plot contours for 
the velocity component along four different directions.
These figures would correspond to velocity-maps if the galaxies were observed 
with the line of sight parallel to these directions. 
The numbers on the contours are the actual velocities along the orbits; an 
observer would measure all velocities multiplied by a factor $\sin i$, with 
$i$ the inclination angle.
The prograde maps (Fig.~\ref{fig:velmapproA}) show clear signs of the 
elongation of the outer orbits, but the overall asymmetries are mild and 
affect mainly the outer parts.

\begin{figure*}
  \centerline{\psfig{figure=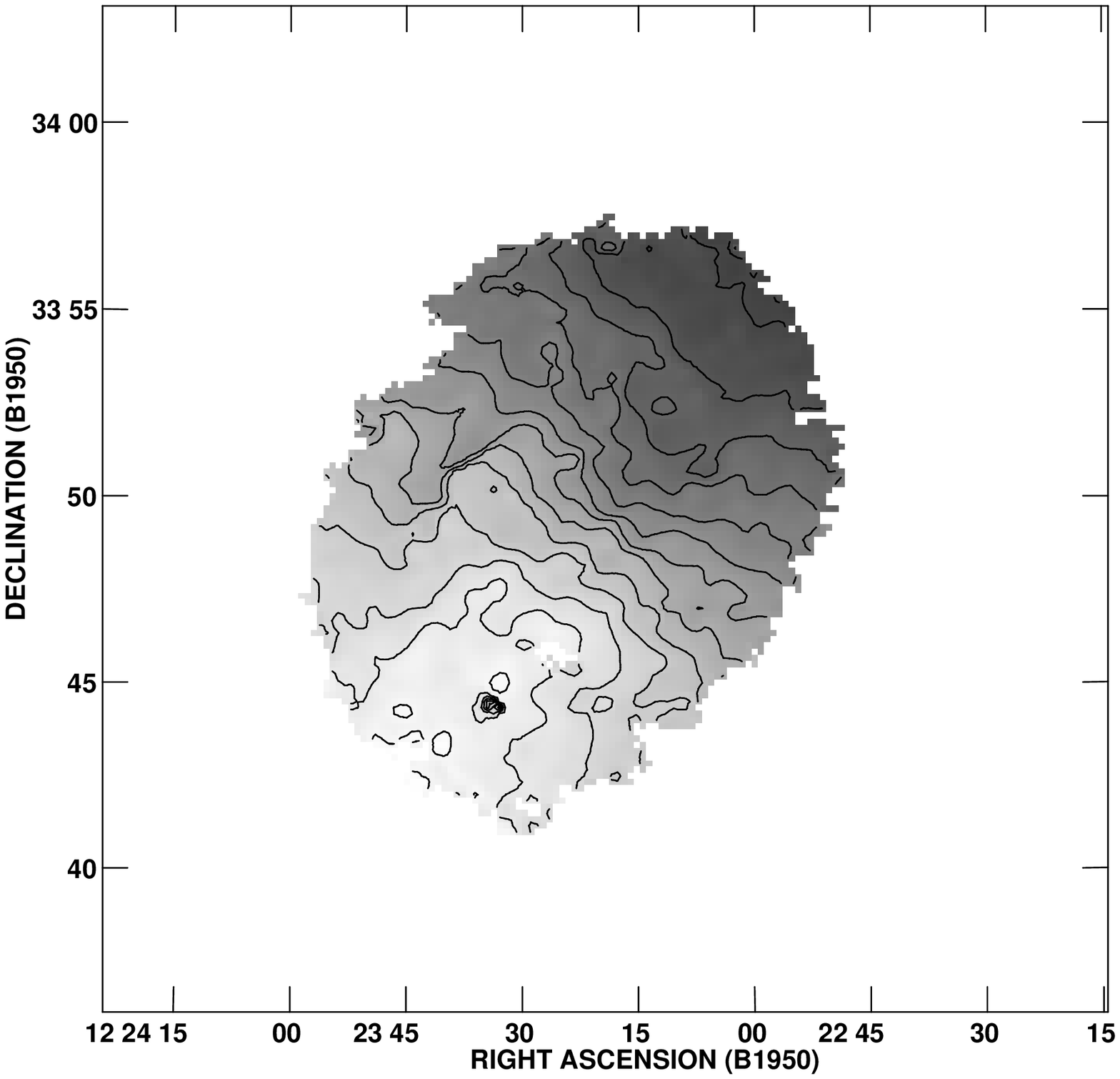,height=4in,angle=-90}}
  \caption{Velocity map from Gaussian fits to \hone\ line profiles at each 
    point in the lopsided galaxy NGC~4395. Dark shading indicates the receding 
    side, contour levels are $260$ to $380 {\rm \, km \, s^{-1}}$, in steps of 
    $10 {\rm \, km \, s^{-1}}$. Map kindly provided by Rob Swaters, 
    from Swaters et al.\shortcite{SwatersMN}.}
  \label{fig:NGC4395}
\end{figure*}

The retrograde maps (Fig.~\ref{fig:velmapretroA}) however, show strong
signs of lopsidedness.  Except when the line of sight is parallel to
the \x-axis (bottom right), the contours in these maps are quite
straight on one side, and strongly curved outwards on the other side.
In Fig.~\ref{fig:NGC4395}, we show the velocity-field of the 
lopsided galaxy NGC~4395, as presented by Swaters et al.\shortcite{SwatersMN}. 
The similarity between this velocity-field and the two upper panels in our 
Fig.~\ref{fig:velmapretroA} is striking. 

From Figures~\ref{fig:orb+curA} to \ref{fig:velmapretroA}, we conclude that 
asymmetries in the overall galaxy potential do not necessarily produce strong 
signals in the rotation curves or velocity fields; depending on the viewing 
angle, the lopsidedness can be hidden.
Conversely, a symmetric rotation curve or velocity-field does not guarantee a 
near-symmetric potential. 

\subsection{Changing the distance from the disc to the halo centre}
\label{subsec:changing dx}

In the simulations by \LS, the spinning disc continued to sink 
towards the halo centre, as long as it was outside the core radius. 
The prograde disc sank further inwards than the retrograde one. 
To investigate whether the orbital structure of our model could explain why
the disc does not stay far outside the halo core, we changed 
the separation $\dx$ between the disc and halo centres,
keeping all other parameters the same as in model A.

\subsubsection*{Model B: $\dx = 1.5$} 
\label{subsubsec:Model B}

\begin{figure*}
  \epsfbox{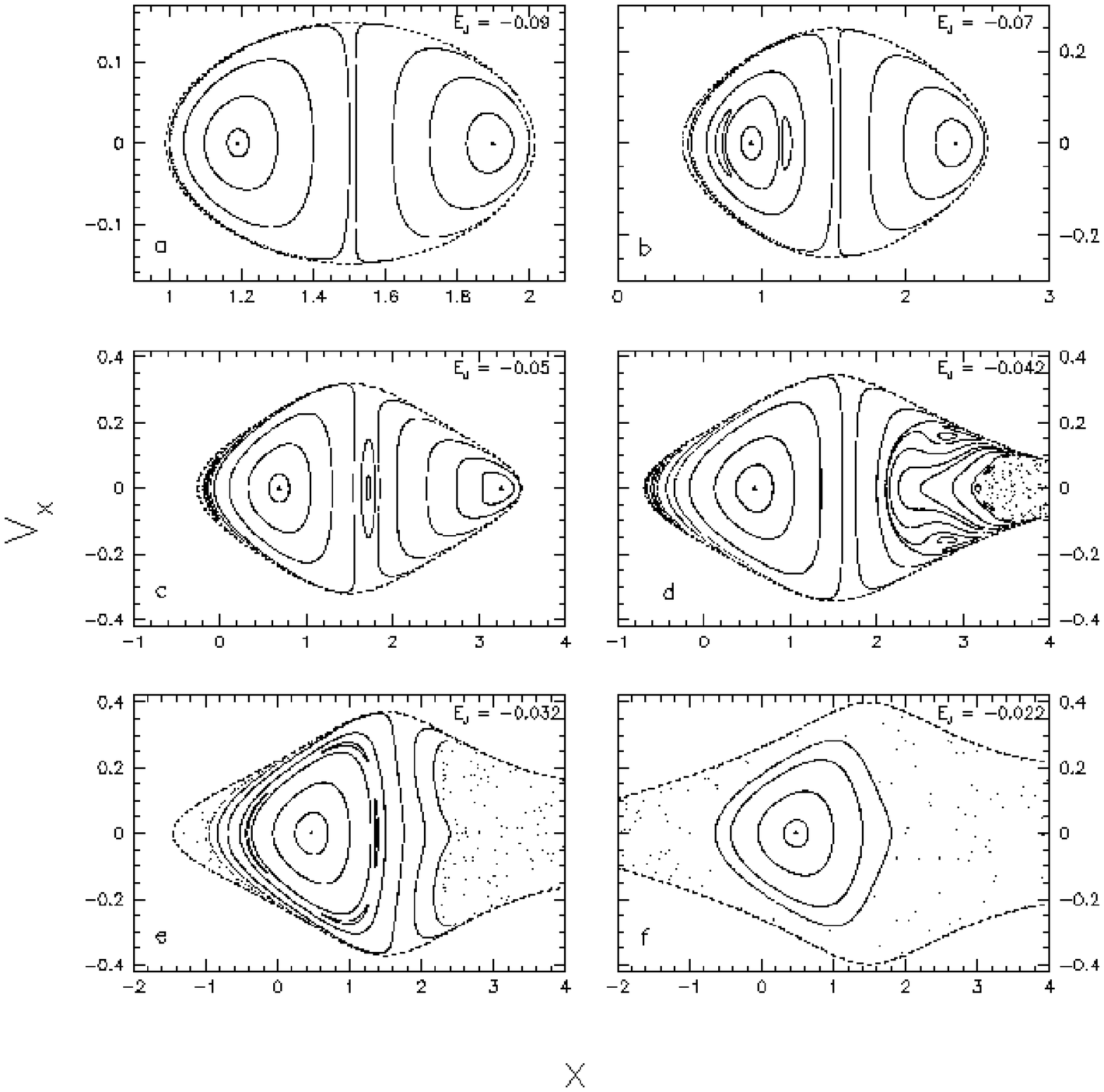}
  \caption{Surfaces of section for $y=0$ and $\vy\geq0$, as in 
   Fig.~\ref{fig:surfsecsA}, but now for model~B\@. The disc centre now lies at 
   $x=1.5$.}
  \label{fig:surfsecsB}
\end{figure*}

In this model, the disc lies inside the halo core, so the halo now contributes 
a larger fraction, 25\%, of the total mass within $r=2$ from the disc centre.
In Fig.~\ref{fig:surfsecsB}, we show the surfaces of section. 
We chose the values for \EJ\ such that the zero velocity curves in each panel 
have approximately the same size and shape as in the respective panels in 
Fig.~\ref{fig:surfsecsA}.
At low values for the Jacobi integral, the surfaces of section show 
qualitatively the same behavior as in model~A; there are two orbit families, 
one prograde (on the right of the surfaces of section) and one retrograde 
(on the left), and almost all orbits are regular. 
As the Jacobi integral is increased, the closed prograde orbit again shifts 
to the edge of the surface of section, while the closed retrograde orbit 
remains in the centre of its family.
Compared to model~A, orbits trapped around the closed prograde orbit continue 
to occupy more of the phase space on the right side of the figures, almost 
until the zero-velocity surface becomes open to the right of the figure.
There are only very few islands of multi-periodic orbits and even at larger 
\EJ, the only irregular orbits are the ones that are not bound to the disc. 
This probably explains why also the prograde disc in \LS\ becomes stable within
the halo core radius, and does not sink much further into the core. 
The retrograde side again remains regular up till high values of \EJ. 

\begin{figure*}
  \epsfbox{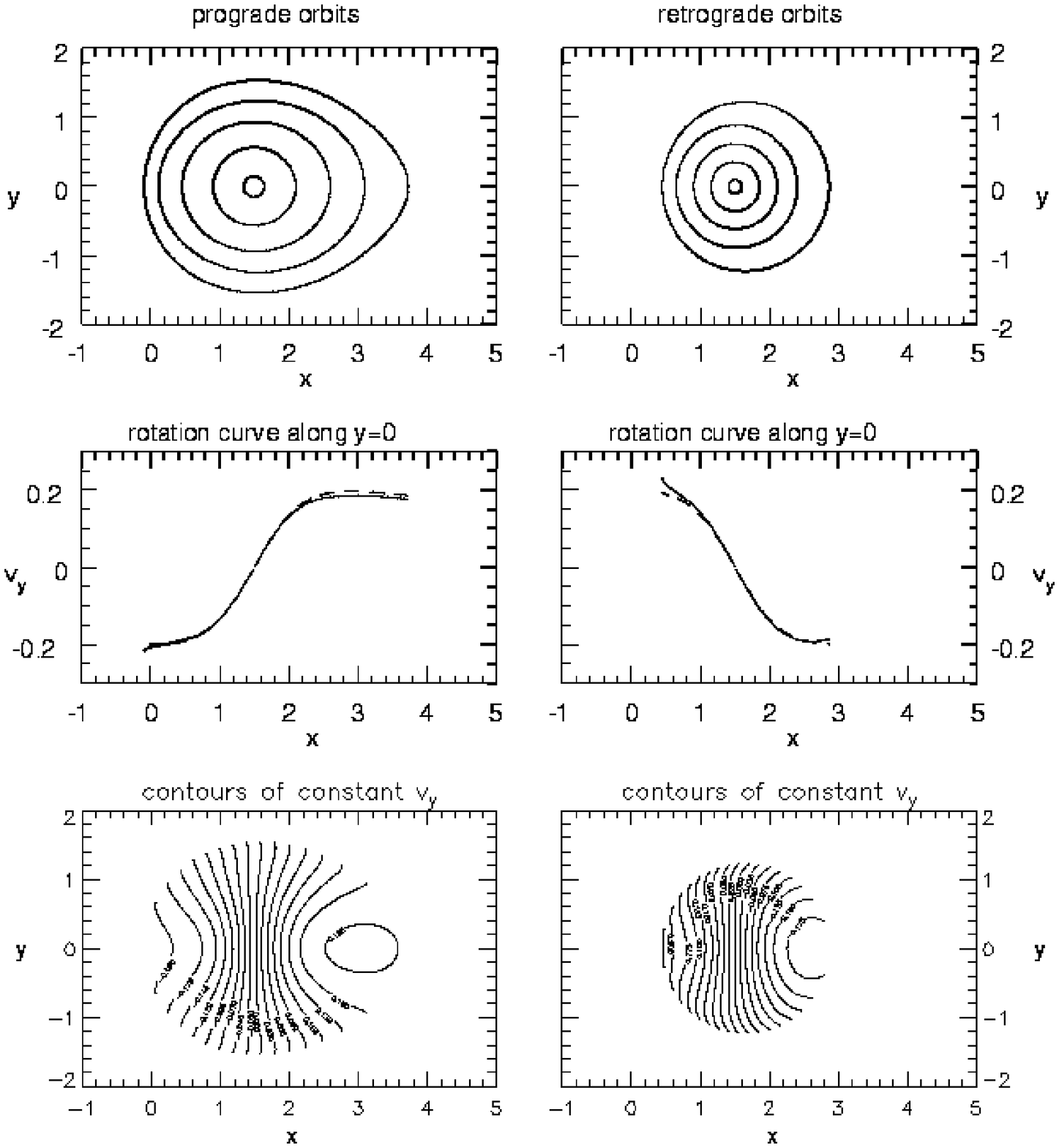}
  \caption{As in Fig.\ref{fig:orb+curA}, the upper panels show closed
    orbits in the prograde and retrograde disks, and middle panels show
    the rotation curves along the \x-axis, but for model~B, with the disc 
    centre at $(1.5, 0)$.
    The bottom panels show contours of constant $\vy$, corresponding to the 
    upper left panels in Fig.~\ref{fig:velmapproA} and \ref{fig:velmapretroA}.}
  \label{fig:orbcurmapB}
\end{figure*}

In Fig.~\ref{fig:orbcurmapB}, we show the closed orbits together with the
rotation curves along the \x-axis and the contours of $\vy$, which
show the most pronounced asymmetries. 
The prograde orbits are now less elongated than in model~A, and their rotation 
curve lies close to the curve of the isolated disc. 
Note that in this case, the prograde rotation curve along the \x-axis also 
becomes flat on the side furthest from the halo centre, and keeps rising on the
other side. 
The model velocity-map of \vy is fairly symmetric.
The rotation curve along the minor axis (at $x=1.5$) rises only
slightly above the curve for an isolated disc. 

The retrograde orbits are slightly more lopsided than in model~A, and the 
asymmetry in the rotation curve along the \x-axis and in the velocity-map of 
$\vy$, is more distinct. 

\subsubsection*{Model C: $\dx = 5$}
\label{subsubsec:Model C}

In model~C we put the disc at $\dx = 5$, far outside the halo core.
This model shows similar behavior to models~D and G, and we refer the
reader to Fig.~\ref{fig:surfsecsD} for an
impression of the surfaces of section.
The prograde side shows highly irregular 
behavior, even well before the zero-velocity curve becomes open.
Large areas in the surfaces of section are occupied by multiply-periodic 
and irregular orbits. 
The region occupied by regular orbits trapped around the closed retrograde 
orbit also starts shrinking at relatively low values of \EJ, when all orbits 
are still bound to the disc.
The irregularity of the surfaces of section probably explains why \LS\ found 
that even the retrograde disc is not stable when far out of the halo centre.
Too few of their simulation particles could follow orbits trapped around
the simple closed orbits.

The rotation curves and velocity-maps are like those of
Fig.~\ref{fig:orbcurmapD}.  The prograde closed orbits become highly
elongated along the \x-axis at large radii, showing high velocity 
gradients where the orbits change from circular to elongated.
Overall, they are fairly symmetric.
The retrograde orbits are only slightly lopsided, and the asymmetries in the
rotation curves and velocity-maps are mild. 
\newline

We conclude that changing the offset of the disc from the halo centre 
affects the prograde orbits most strongly. 
As we increase the offset, the outer orbits become highly elongated, and 
high velocity gradients develop where the orbits change 
from circular to elongated. 
The prograde side of the surfaces of section becomes complicated, with few 
regular orbits trapped around the closed orbits.  
Hence a prograde disc is unlikely to survive unless it is well within the 
halo core.

The retrograde orbits and rotation curves are less affected, and
become only slightly less asymmetric as we increase the offset.   
At low separations, the velocity map of \vy is highly asymmetric,
while increasing the offset makes the asymmetry less pronounced.
As the disc is moved away from the halo core, fewer retrograde orbits
are trapped around the closed periodic orbit.

\subsection{Changing the disc mass}
\label{subsec:changing massratio}

Changing the mass of the disc in our models has similar effects on the
kinematics as changing the offset of the disc from the halo centre.
We expect a model with a heavy disc to produce qualitatively the same
behavior as a model with the disc far out from the halo centre.
Indeed, the strength of the asymmetries in our model kinematics
appears closely related to the relative densities of the disc and the
halo, that is, to the parameter $\eta$.

\subsubsection*{Model D: $\Md = 0.186$}
\label{subsubsec:Model D}

In model~D, the disc is twice as heavy as in the previous models. 
The offset between disc and halo centre is the same as in model~A: 
$\dx = 2.5$.
The halo now has a much smaller influence on the inner parts of the
disc, with $\eta = 9\%$. 
The potential well of the disc is much deeper now, so we 
need to examine lower values for \EJ\ to see similar orbits. 
\begin{figure*}
  \epsfbox{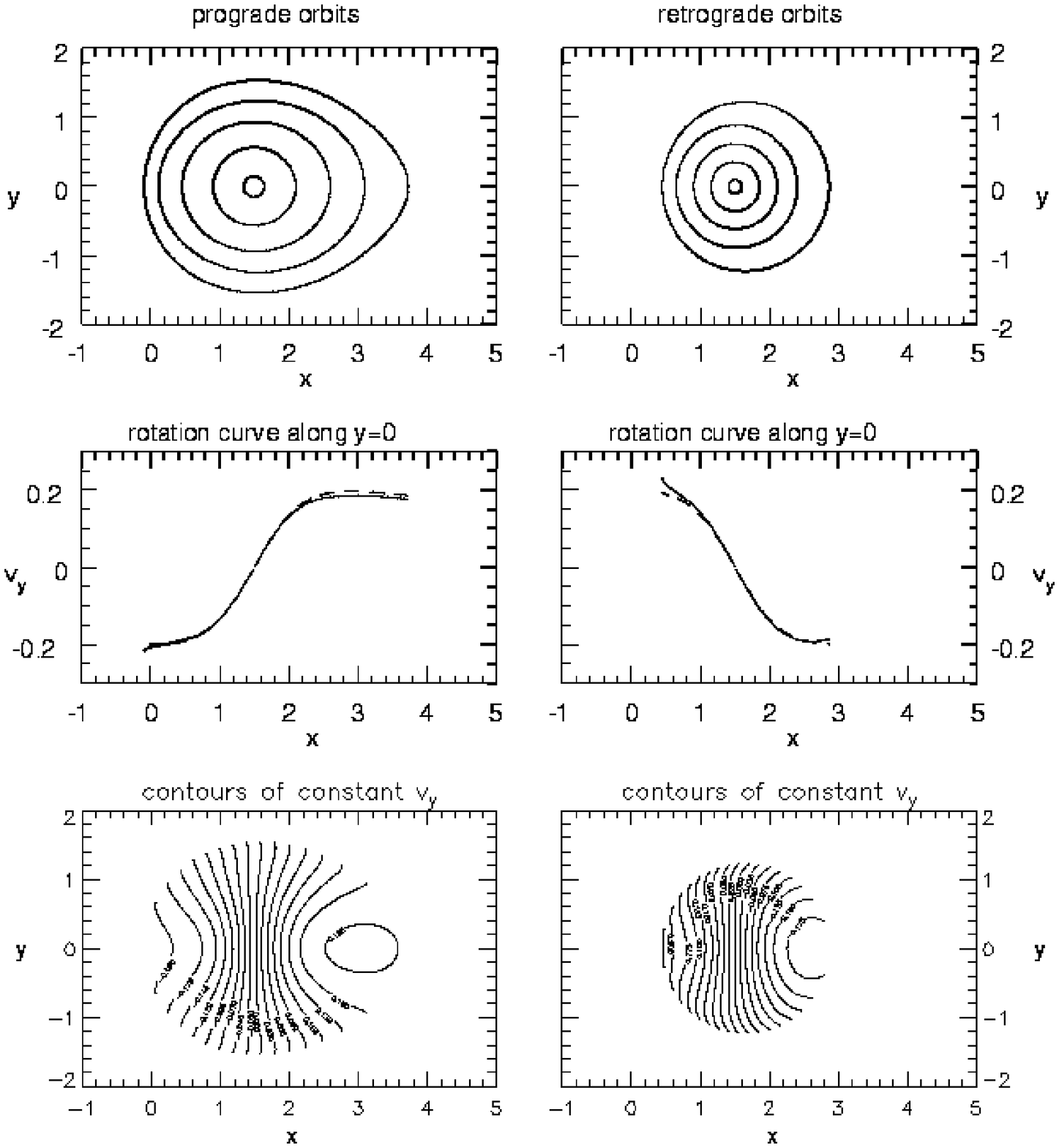}
  \caption{Surfaces of section for $y=0$ and $\vy\geq0$, as in
    Fig.~\ref{fig:surfsecsA}, but now for model~D\@. The disc centre 
    lies again at $x=2.5$, but the disc mass is $\Md = 0.186$.}
  \label{fig:surfsecsD}
\end{figure*}
Again, most retrograde orbits are regular up till high values for \EJ. 
The prograde side of the plots becomes distorted, with a large irregular 
region starting from $\EJ = -0.09$. 
For a limited range in \EJ, there is an additional prograde family 
of closed orbits, visible in Fig.~\ref{fig:surfsecsD}c near $(3.8,0)$. 
Orbits trapped around these closed orbits occupy only a small region in phase 
space and are not expected to play an important role in any realistic
model.

The same reasoning that we used in discussing model~C leads 
us to suspect that a massive prograde disc will not be stable,
and that it will sink towards the halo centre rapidly. 
By contrast, a retrograde disc might be stable when orbiting just 
outside the halo core.

\begin{figure*}
  \epsfbox{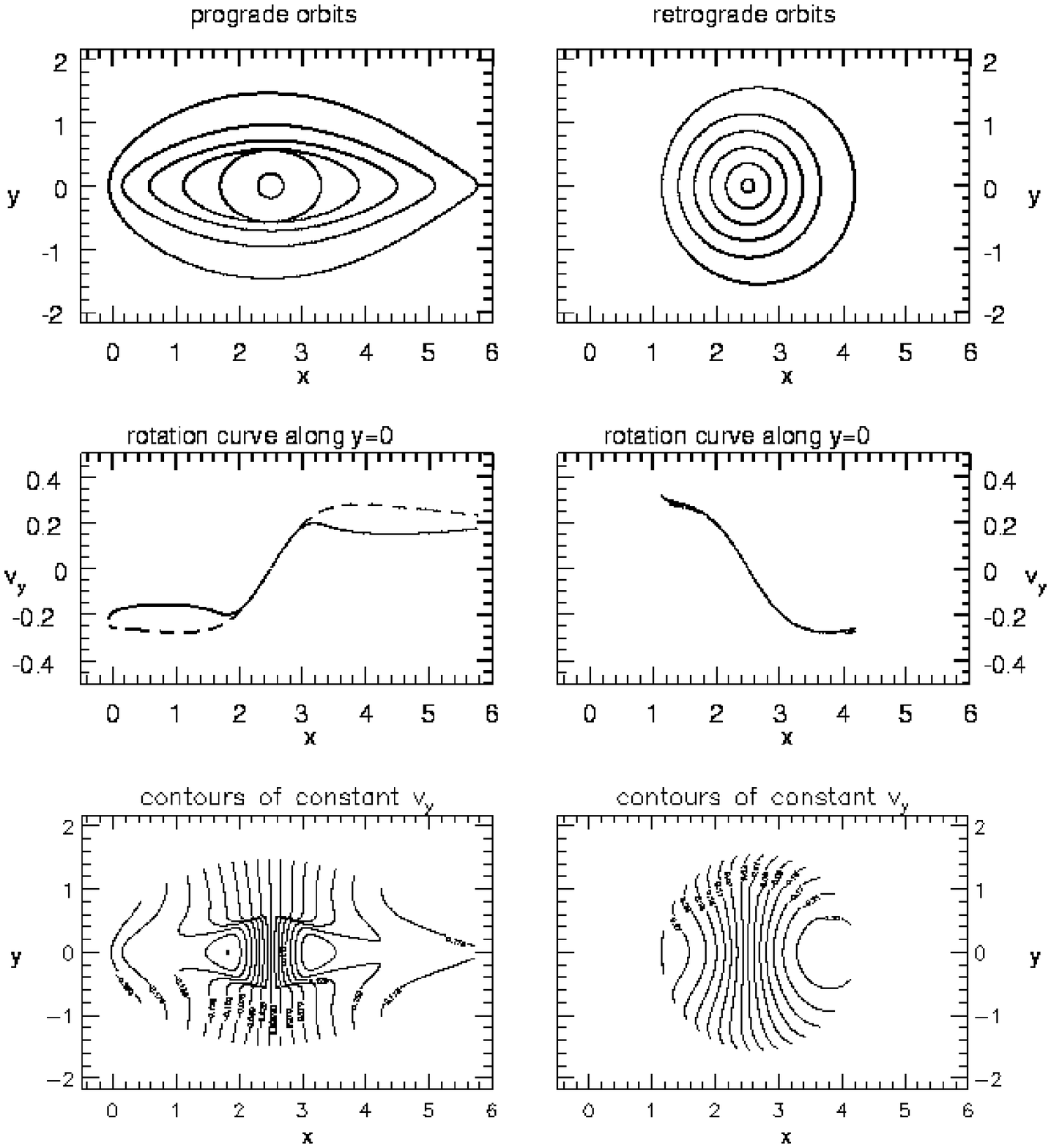}
  \caption{As in Fig.~\ref{fig:orbcurmapB}, the upper panels show closed
    orbits in the prograde and retrograde disks, middle panels show
    the rotation curves along the \x-axis, and lower panels contours of 
    constant $\vy$, but now for model~D\@. The disc centre lies at $x=2.5$, 
    and the disc mass is $\Md = 0.186$.}
  \label{fig:orbcurmapD}
\end{figure*}
In Fig.~\ref{fig:orbcurmapD}, we show the orbits, the rotation curves along the
\x-axis and the contours of \vy in this model. 
Non-crossing periodic orbits extend to larger distances from the disc
centre than they did in model~A.
The prograde orbits become highly elongated at larger radii, and their
major-axis rotation curve lies far below the curve for an isolated
disc.  The contours of \vy show strong velocity gradients in this
region, but are overall quite symmetric. 
The rotation curve along the minor axis is symmetric,
but shows a large, abrupt rise in velocity at the radius where the orbits 
change from circular to elongated. 

The retrograde orbits look similar to the ones in the previous models. The
asymmetries in the rotation curve along the \x-axis, 
and in the velocity-map of $\vy$, are less severe than in models~A and B.

\subsubsection*{Model E: $\Md = 0.046$}
\label{subsubsec:Model E}

Model~E, with a low-mass disc of $\Md = 0.046$, shows many
similarities with our model~B.  Its surfaces of section show few
irregular orbits, with many orbits trapped around the closed orbits.
Because of the regularity of the surfaces of section, we expect that a
low-mass disc with either prograde or retrograde spin will be stable
near the halo core radius in N-body simulations.

The closed prograde orbits are only slightly elongated; the rotation 
curve along the \x-axis becomes flat on the side more distant from the halo
centre, and keeps rising on the other.
The velocity-maps are only mildly asymmetric.
The rotation curve along the \x-axis for the retrograde orbits is again 
clearly lopsided, and the velocity-maps are generally highly asymmetric.

\subsection{Changing the halo core radius}
\label{subsec:changing halosize}

\LS\ noted that when the halo core radius was small ($\rc = 0.5$),
both the prograde and retrograde disc sank towards the halo centre rapidly. 
Decreasing the core radius of the halo has the effect of 
compressing the material in the halo, and thus of increasing its
central density.

\subsubsection*{Model F: $\rc = 1$}
\label{subsubsec:Model F}

In model~F, the halo core radius is $\rc = 1$, half the size as in model~A 
and equal to the disc scale length. The disc centre lies at $\dx = 1.5$, 
just outside of the halo core. 
The halo now contributes 37\% of the total mass within $r=2$ from the disc 
centre. 

\begin{figure*}
  \epsfbox{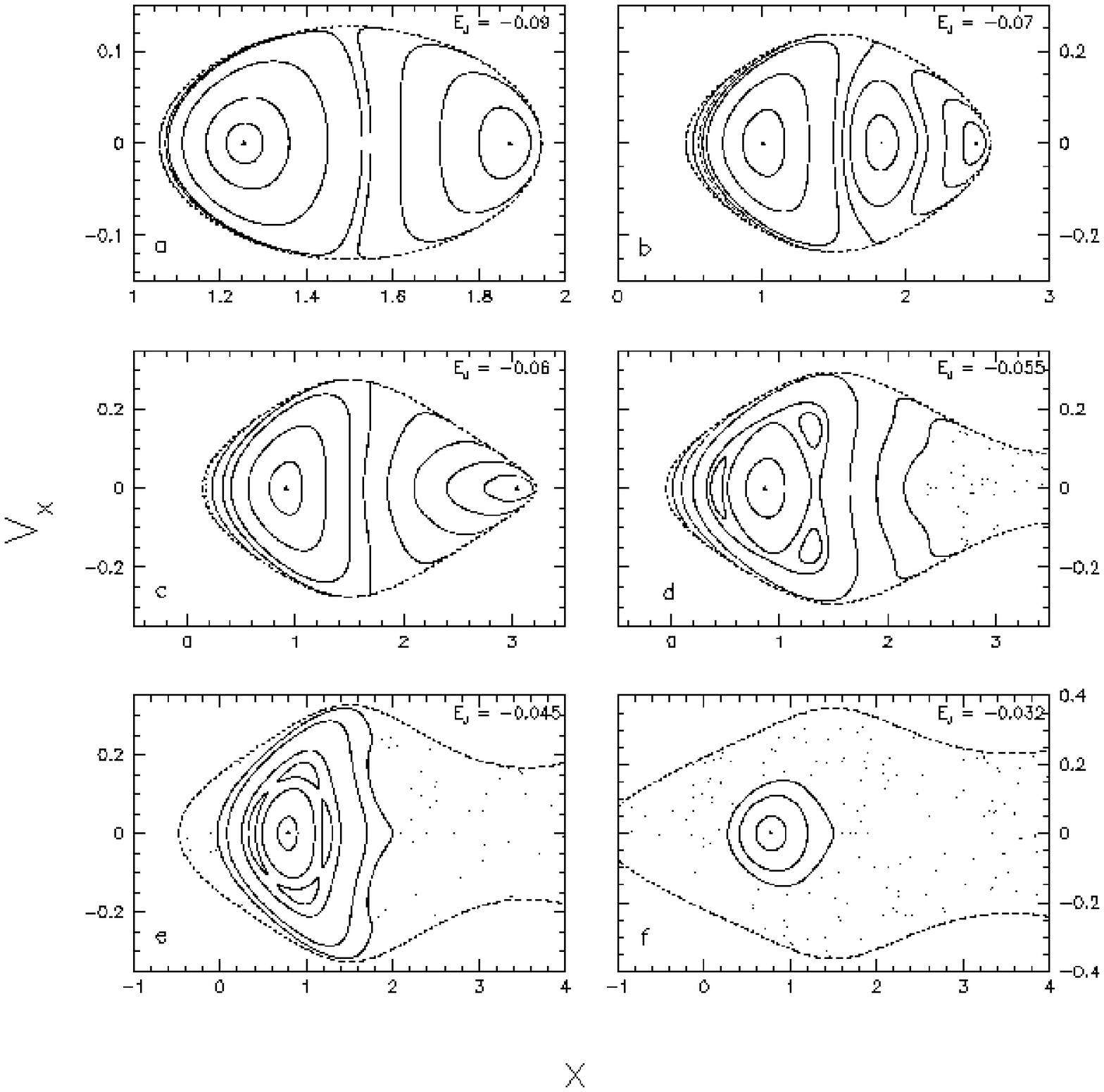}
  \caption{Surfaces of section for $y=0$ and $\vy\geq0$, as in
    Fig.~\ref{fig:surfsecsA}, but now for model~F\@. The halo now has
    a core radius of $\rc = 1$. The disc centre lies at $(1.5, 0)$.}
  \label{fig:surfsecsF}
\end{figure*}

In Fig.~\ref{fig:surfsecsF} we show the surfaces of section. 
There are again very few irregular orbits. 
In panel~a, we recognize the two familiar families of prograde and retrograde 
orbits. 
In panel~b, the original prograde family has shifted to the right, and a large 
portion of the prograde side, around $x=1.8$, is now occupied by a
third family of orbits. 
These orbits have a loop that appears as an additional, narrow island in the 
far left corner of this panel, near $(0.5, 0)$. 
At $\EJ = -0.06$ (panel~c), the original prograde family has disappeared; 
meanwhile, the third orbit family has lost its loop on the left side of the 
surface of section, and has now become simple-periodic itself. 
Because these orbits are simple-periodic only for a limited range in
the Jacobi integral, 
within which they cross each other abundantly, we do  
not use them to make model rotation curves or velocity-maps. 
However, the fact that such a big part of phase-space is occupied by this 
family gives an indication why the prograde disc was not stable in the 
simulations by \LS. 
Only at low values for the Jacobi integral can many stars be on orbits trapped 
around the closed prograde orbit; as the Jacobi integral is increased, stars 
follow different orbits, resulting in strong mixing in the disc. 

The retrograde side of the surfaces of section remains regular up to high 
values of \EJ.  The range 
where we see closed retrograde orbits ($ -0.1 \la \EJ \la 
-0.03$) is about twice as large as the range where the familiar closed 
prograde orbits exist ($-0.1 \la \EJ \la -0.065$). 
This difference is more extreme than in the previous models. 

\begin{figure*}
  \epsfbox{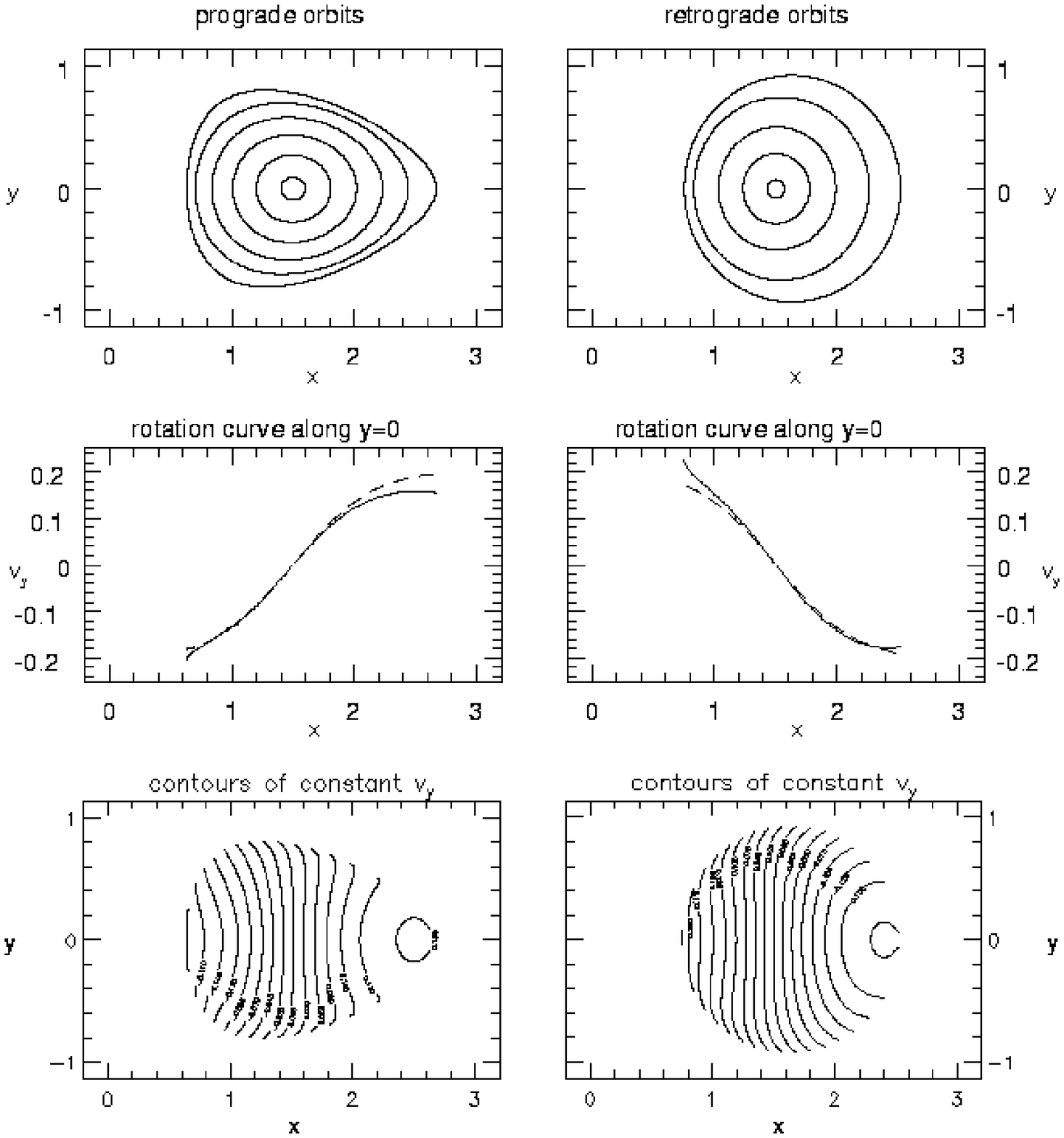}
  \caption{As in Fig.\ref{fig:orbcurmapB}, the upper panels show closed
     orbits in the prograde and retrograde disks, middle panels show
     the rotation curves along the \x-axis, and lower panels 
     contours of constant $\vy$, but now for model~F\@. The disc 
     centre lies at $(1.5, 0)$, and the halo has a core radius of $\rc = 1$ 
     now.}
  \label{fig:orbcurmapF}
\end{figure*}

In Fig.~\ref{fig:orbcurmapF}, we show orbits, rotation curves and
velocity-maps in this model.  
Neither the prograde or the retrograde closed orbits extend more than
one scale length from the disc centre.  This would explain 
why neither the prograde nor the retrograde disc was stable in the
simulations of \LS; only particles in a small central part of the disc
can be trapped near the simple closed orbits. 

Both the prograde and the retrograde rotation curves along the \x-axis
have a flat part on one side and a rising part on the other. 
The appearance is similar to the retrograde orbits in our models~B and
E, but with more extreme asymmetry. 
The minor-axis rotation curves 
deviate only slightly from the curve of an isolated disc. 
The velocity-maps are very asymmetric, especially the retrograde map.

\subsubsection*{Model G: $\rc = 4$}
\label{subsubsec:Model G}

Model~G, with a large halo ($\rc = \dx = 4$), shows similar behaviour
to our models~C and D. The surfaces of section show many irregular orbits on 
the prograde side, while the retrograde side again stays regular till high 
values of \EJ. 
As in model~D, the prograde orbits become highly elongated along the \x-axis;
the rotation curves and velocity-maps look like the ones in 
Fig.~\ref{fig:orbcurmapD}, with high velocity gradients where the orbits 
change from circular to elongated. 
The retrograde orbits are only slightly lopsided, and the rotation curves and 
velocity-maps are hardly asymmetric. \newline

The analysis of models with different sizes of the halo confirms our
conjecture that the shape of the orbits and the velocity contours
depends mainly on the relative densities of the disc and 
the halo, as given by the parameter $\eta$.
In model~F, with a compact halo, 
the rotation curves and velocity-maps resemble those of
models~B and E, which both have a relatively high halo density too.
The rotation curves along the \x-axis show a flat part on the far side
from the halo core, and a rising side on the other side, for both the
prograde and retrograde orbits.  
The retrograde velocity-maps of \vy have strongly curved contours on
the right side, but almost completely straight contours on the other side.

Model~G, with a large, extended halo, shows
the same behavior as models~C and D, which also have a low halo density.
The prograde side of the surfaces of section has
large irregular regions, and the closed orbits become highly elongated
at large radii, causing strong velocity gradients in the rotation
curves and velocity-maps. The retrograde orbits, rotation curves and
velocity-maps are only mildly asymmetric.

\section{Discussion and conclusions}
\label{sec:discussion}

We have investigated an analytical model for lopsided galaxies, based on 
numerical simulations by Levine \& Sparke \shortcite{LS}, in which a disc 
orbits off-centre within a `dark matter' halo. 
We have constructed orbits, rotation curves and velocity-maps in the resulting 
gravitational potential; all these show significant asymmetries. 
The asymmetries show a clear correlation with the relative densities
of the halo and the disc; they are stronger in the models where the
halo contributes more of the mass in the inner regions of the disc
(models~B, E and F).

In the retrograde case, the orbits, and hence the expected
distribution of stars or gas, are sometimes only slightly lopsided;
but the rotation curves and velocity maps can be highly asymmetric.
The rotation curves along the \x-axis, the symmetry axis of the
lopsided system, {\it keep rising} on the side near the halo centre
but {\it become flat} on the outer side.  Correspondingly, velocity 
contours are straight on the side near the halo centre and curved 
on the outer side.   
Generically, the peak of the global \hone-profile on the
side where the rotation curve is flat will be higher, while the side
with the  lower peak has a sloping `shoulder' to high velocity.
This characteristic asymmetry is similar to that in the velocity-map
of the lopsided Sm galaxy NGC~4395 presented by Swaters et
al. \shortcite{SwatersMN}. 
This late-type system is one of the lowest-luminosity galaxies to
have a Seyfert nucleus, which in turn is the faintest Seyfert~1
nucleus known  \cite{FS}, with $M_B = -10.3$ or 
$L = 2 \times 10^6 \lsun$ \cite{Matthews2}.  
Presumably a central black hole is present in the nucleus.

The kinematics of our models 
are not unique but a general feature, in the sense that different
combinations of parameters can give similar features in  the rotation
curves or velocity-maps.  We can use the strength of the asymmetries  
to determine the parameter $\eta$, the
fraction of mass contributed by the halo within 2 scale lengths of the
disc centre.
For example, the velocity field of the galaxy NGC~4395 looks similar
to some of our retrograde models.  Comparing our
Fig.~\ref{fig:NGC4395} to the velocity-maps of models~A, B, E and F,
we estimate that this galaxy has a value for $\eta$ between 20 and
30\%.  The rotation curves of these models lie always close to the
curves of the isolated disc (in other words, these models are `maximum
disc'); we can use the total width of the rotation curve 
($v_{{\mit max}} - v_{{\mit min}}$) as a measure of the
disc mass.  We infer a total mass for the disc of NGC~4395 of about
$3.5 \cdot 10^{9} {\rm M}_{\odot}$ within 8.5 kpc of the centre,
taking Swaters' distance of 3.8 Mpc and inclination $i=46^{\circ}$.
The blue magnitude of Matthews et al. \shortcite{Matthews2} implies a
mass-to-light ratio $M/L_B \approx 2$, which is reasonable for the
stellar population of a late-type disc.

Although the simulations of \LS\ showed that prograde-spinning discs
sank more rapidly into the halo core than retrograde ones, we may
still expect to observe some lopsided prograde discs as a transient
phenomenon.  In these, the gas distribution has a pronounced
egg-shaped appearance; but the rotation curve can be very nearly the
same on both sides of the galaxy, as is observed in some real galaxies.  
In models where the halo density is
lower (models~C, D and G), these prograde orbits become highly
elongated, and strong velocity gradients develop.  These orbits are
probably not realistic for gas moving in such a galaxy.

It is interesting to compare our velocity fields to the ones found by
Schoenmakers \shortcite[chapter~6]{Schoenmakers}. He used epicycle
theory to derive velocity fields in a model with an offset between the
disc and halo centres.  The mass ratio of his model is comparable to
our model~E ($\Md \approx 0.05$), but the offset is much smaller ($\dx
\approx 0.8 \Rc$). His velocity fields are less asymmetric than ours, even 
though the halo density in the inner parts of his model is higher than in 
any of our models. This difference is probably caused by the fact that 
Schoenmakers assumed that the offset between the disc and halo remained 
static in an inertial frame. Also, because of the small offset, the region 
where we expect to see a characteristic distortion of the velocity field is 
very small.     

Our models explain the findings of Odewahn \shortcite{Odewahn} and
Matthews et al. \shortcite{Matthews1} that lopsidedness is more
common in discs of extreme late type (Sd and later). 
Broeils \shortcite[Figure 9b of Chapter~10]{Broeils} shows that the
ratio of the dark halo's core radius to the optical radius
$R_{25}$ is $\geq 1$ for all but one of the systems of type Sd or
later, while showing a large scatter for earlier types. Since our
models show the most pronounced lopsidedness when the halo core is
larger then the disk scale length, we expect most late-type disks, but
a smaller fraction of the earlier types, to be highly susceptible to
lopsidedness. Just as in the models of Jog \shortcite{Jogone,Jogtwo},
we expect long-lived lopsidedness to be more severe, and thus more
easily detected, in systems where the dark halo dominates the mass
distribution.

We have calculated the closed orbits within the disk only until they
begin to cross each other, or are no longer closed.  These orbits 
are usually found only within $\sim 2$ scale lengths of the disk center.  
This is roughly the region probed by the optical studies of
Zaritsky \& Rix \shortcite{ZR} and Rudnick \& Rix \shortcite{RR}, 
but does not extend as far out as some of the \hone\ studies of
giant disk galaxies. Gas and stars may still circle the disc center
outside our orbits; we would still expect lopsided kinematics there.
Also, the disk would lie off-center with respect to any gas on orbits at
larger distances that circle the halo center; but we do not make any
detailed predictions for these cases. 

We do not know whether an off-center disk can account for {\it all}
types of lopsidedness.  The best test is to examine both the velocity
field and the distribution of stars and gas in the same
galaxies.  Neither Swaters \shortcite{Swatersthesis} nor Kornreich et
al.\ \shortcite{Kornreich1,Kornreich2} found any appreciable
correlation between morphological and kinematical lopsidedness.  
The models of Syer \& Tremaine \shortcite{ST}, Earn \& Lynden-Bell 
\shortcite{ELB} and Jog \shortcite{Jogtwo}, in which the centers of the disk 
and halo remain coincident, predict that photometrically-lopsided disks
should also be kinematically lopsided.  Our model can produce either
photometric or kinematic lopsidedness, with very little of the other.

In all our models, the rotation curves perpendicular to the direction
of lopsidedness (the \x-axis) are symmetric, and the velocity-maps of
\vx are only mildly asymmetric.  
Thus even a quite lopsided galaxy can look symmetric when seen from an 
unfavorable direction.  In the retrograde models, velocity asymmetries
are apparent for a fairly large range in viewing angles; an offset of 
$30^{\circ}$ from the \x-axis is usually enough.
Kinematic asymmetries in the prograde discs are only visible for a 
small range of viewing angles, within  $30^{\circ}$ of the most 
favorable direction.  Naturally, the asymmetry in the prograde orbits 
themselves can be seen from any direction.
The fact that lopsidedness can be hidden by an unfavorable viewing
angle has important implications for the frequency with which it
actually occurs in nature.  The actual fraction of all galaxies that
is lopsided may be higher than the 30--50\% inferred from kinematic 
observations such as those of Richter \& Sancisi \shortcite{RS}, 
Haynes et al. \shortcite{Haynes} and Matthews et al. \shortcite{Matthews1}.

The rotation curves for the prograde models often deviate far from the curves
for an isolated disc of the same mass, even though they are usually quite 
symmetric (these models are {\it not} maximum disc); 
For example, in model~A, where the prograde orbits are still
well-behaved, the actual maximum in the rotation curve differs up to 
about 15\% from that for the isolated disc, depending on the viewing angle.  
This effect will
lead to incorrect estimates for the disc mass, and to scatter in the
Tully-Fisher relation between rotation speed and luminosity of disc
galaxies.

What might be the origin of lopsidedness? Generically, galaxies are
widely believed to form by repeated mergers at early times, while much
of the disk material may represent late infall.
Walker, Mihos and Hernquist \shortcite{Walker}, and Z\&R \shortcite{ZR}, 
proposed that merging a small galaxy with a larger disk system should produce 
a lopsided disk.  If the fractions of dark and luminous
matter are not the same for all fragments, we may expect the center of
the disk to be displaced from that of the dark halo.  As an example,
if the small system is largely gaseous, and comes in with nonzero
angular momentum, then once it has merged with the larger galaxy's
luminous disk, the center of the combined disk will orbit that of the
halo.

Although the agreement of our model velocity fields with observations
is promising, the stability of our models remains an important issue.
For example, we have modelled the halo as a fixed potential, ignoring 
dynamical friction between the disc and whatever particles make up the 
unseen halo. 
However, it is not clear that dynamical friction must cause 
the lopsidedness to damp rapidly.
Weinberg \shortcite{Weinberg1,Weinberg2} and Vesperini \& Weinberg 
\shortcite{VW} found that some spherical galaxy models have
long-lived lopsided modes; once excited, these modes take many
orbital times to decay.  Taga \& Iye \shortcite{TI} have shown that a central
massive object can wander away from the centre of an initially
spherical stellar system.  They found instability 
only when the central object had no more than 10\% of the total mass;
but Weinberg's work implies that somewhat larger masses should suffer
only weak damping.  These indications are promising, but further
calculations will be needed to explore the stability of our models.

\section*{Acknowledgments}

We are grateful to R.A. Swaters for helpful discussions about
observations of lopsided galaxies, and to our referee, whose reports
helped us to clarify our presentation. E.N. wants to thank the
Astronomy Department of UW-Madison for hospitality during the course
of this work, and the Kapteyn-fonds and Utrecht University for
financial support. We thank the US National Science Foundation for
support under grant AST-9803114.

\end{document}